\documentclass[12pt]{iopart}
\usepackage[dvips]{graphicx}

\begin{document}

\title{Partial scaling transform of multiqubit states as a criterion of separability}

\author{C. Lupo$^{1}$, V. Man'ko$^{2}$, G. Marmo$^{1}$ and E. C. G. Sudarshan$^{3}$}

\address{$^1$ Dipartimento di Fisica dell'Universit\`a di Napoli Federico II
and Istituto Nazionale di Fisica Nucleare (INFN) Sezione di Napoli,
Complesso Universitario di Monte Sant'Angelo, via Cinthia, Napoli, I-80126, Italy}
\address{$^2$ P.N.Lebedev Physical Institute, Leninskii Prospect 53, Moscow 119991, Russia}
\address{$^3$ University of Texas, Center for Statistical Mechanics, 1 University Station C1609, Austin, TX, USA}
\eads{\mailto{lupo@na.infn.it}, 
\mailto{manko@sci.lebedev.ru}, 
\mailto{marmo@na.infn.it}, 
\mailto{sudarshan@physics.utexas.edu}}

\begin{abstract}

The partial scaling transform of the density matrix for multiqubit states is introduced to detect entanglement
of the quantum state.
The transform contains partial transposition as a special case.
The scaling transform corresponds to partial time scaling of subsystem (or partial Planck's constant scaling)
which was used to formulate recently separability criterion for continous variables.
A measure of entanglement which is a generalization of negativity measure is introduced being based on
tomographic probability description of spin states.

\end{abstract}

\maketitle

\section{Introduction}

Entanglement is a purely quantum characteristic of composite system \cite{Schr} and it creates quantum correlations
among the subsystems of the system.
Till now there is no complete understanding and protocols to detect entanglement for generic multipartite system both
for qubit (qu$d$it) case and for continous variables like photon quadrature components in the multimode electromagnetic
field.
There exists partial positive transpose criterion of separability \cite{Peres,Horod,Simon}, this criterion gives
necessary condition for separability of composed bi-partite system.
In the case of two qubits and qubit-qutrit system this criterion is also a sufficient condition for separability
which allows to detect entanglement without ambiguity.
For all the other systems of qudits the positivity of partial transpose gives the necessary condition for separability
only.
The partial positive transpose (PPT) was also shown \cite{Simon} to provide the necessary and sufficient condition
of separability for a pure two-mode Gaussian state of photons.
From the point of view of physics the partial transpose of subsystem density matrix means partial time reversal for
the subsystem constituents.
Obviously the change of time sign for all the system is an admissible operation in the sense that the density matrix
obtained as result of time reversal corresponds to a state that can be realised in nature.
But in case of entangled state change of sign of time for a subsystem constituents only is not an innocent operation
and it can provide as result the new operator from the initial density operator of the system which is not an
admissible density operator.
The obtained operator can have negative eigenvalues which make it unappropriate to serve as a density operator.
The transformations of density operator which provide as result a hermitian positive trace-class operator
(which can serve as density operator) form the set of positive maps \cite{Choi,Voronow} or dynamical maps \cite{Sudar61,nonCP}.
The transpose transform is an example of positive maps.
The positive map of initial state density operator can be realised by time evolution of the system: it is just a dynamical
map.
Recently it was pointed out \cite{Olga,Anil} that there exist positive maps which are related to other transforms,
e.g. scaling transforms of time, Planck's constant and other universal constants.
In \cite{Anil} the scaling of time (or scaling of Planck's constant) is applied to a subsystem of a composed system
to formulate a new criterion of separability for multimode photon states.
This criterion extends the criterion based on partial transpose operation and contains continous parameters
describing the positive map of density operators.
For entangled states the scaling provides as result operators with negative eigenvalues.
Due to this the partial scaling transform can serve as necessary and sufficient condition of separability of
Gaussian states of multimode electromagnetic field which extends result of \cite{Simon} for bi-partite systems
only.
One can understand the time scaling transform as change of momentum of the particle $p\rightarrow\lambda p$.
Obviously such change of momentum induced the transform of angular momentum (spin) also.

Till now the partial scaling transform of time (or Planck's constant) was not used to detect
entanglement in multiqubit (multiqudit) system.
The aim of this letter is to introduce for the multiqubit system the map of density operator based on partial scaling
transform of time for the constituents of subsystems and to use the transform to distinguish the separable and the
entangled states.
We also apply the description of qubit states by means of probability distribution (tomograms) \cite{OlgaJ}
to consider a measure of entanglement suggested in \cite{SudZac}.

The letter is organised as follows.

In Section \ref{partial transpose} partial transpose of density matrix is discussed.
In Section \ref{Bipartite and tripartite} notion of bipartite and tripartite entanglement is considered.
In Section \ref{Scaling transform} a scaling transformations for qu$d$dits is introduced.
Some examples (Werner \cite{Werner}, GHZ \cite{GHZ} and W states \cite{Wstates}) for two and three qubits and qutrits are studied in Section \ref{examples}. 
In Section \ref{outlook} and Section \ref{conclusions} conclusions and further topics are presented.

\section{\label{partial transpose}Partial transpose}

The operation of taking the transpose of a matrix defines a positive map:
\begin{equation}
T \ \ : \ \ \rho \rightarrow \rho^T \ \ \rho \geq 0 \Rightarrow \rho^T \geq 0
\end{equation}
it is well known that this map is positive but not completely positive: by the Peres-Horodecki \cite{Horod} 
criterion this gives a necessary condition for a state of a bipartite system to be separable.
Physically this map is equivalent to time reversal:
\begin{equation}
t \longrightarrow -t
\end{equation}

Suppose we have a bipartite spin system: $S=S_1 \times S_2$, this yields the decomposition of the Hilbert space 
of the whole system as a tensor product $\mathcal{H}=\mathcal{H}_1 \otimes \mathcal{H}_2$.
A density state for the system $S$ is described by a density matrix $R$, the decomposition of the Hilbert space 
as tensor product induces the matrix representation $R \equiv R_{a\alpha,b\beta}$, where the latin and greek indices 
refer to first and the second system respectively.
One defines the operation of partial transpose that consists in performing the transposition only on the indices of, 
say, the second system:
\begin{equation}
T_2 = I \otimes T \ \ : \ \ R \equiv R_{a\alpha,b\beta} \rightarrow R^{T_2}\equiv R_{a\beta,b\alpha}
\end{equation}
This operation can be seen as a partial time reversal that acts on the second subsystem only: if the states present only
classical correlations the map is always well defined and provides as result an admissible density operator
of composite system.
Otherwise, in presence of quantum entanglement the map may transform a density matrix of composite system into 
a matrix that is not positive.

\section{\label{Bipartite and tripartite}Bipartite and tripartite entanglement}

Positive but not completely positive maps are known to provide entanglement witnesses \cite{Horod}.
The usual approach is to take a bipartite quantum system with Hilbert space of states 
$\mathcal{H}_1 \otimes \mathcal{H}_2$ and apply a map like the following one:
\begin{equation} \label{wit}
\mathcal{P}_2= I \otimes P
\end{equation}
that acts trivially on the first subsystem and by a positive but not completely positive map $P$ on the second one.

Partial transposition is a special example of such a procedure, in this case we take:
\begin{equation} \label{pt}
\mathcal{T}_2 = I \otimes T
\end{equation}
In general for a given $P$ the criterion is only necessary for separability of states.
But for the whole set of positive but not completely positive maps, the criterion becomes also sufficient.
This is not a simple issue because it is not yet known how to characterize positive but not completely positive maps 
for a system that is not a qubit.
Nevertheless it is known that for a two qubit and a qubit-qutrit system \cite{Horod} and for pure Gaussian states \cite{Simon} the criterion 
is also sufficient.

Even for qubit system the situation becomes very different when one deals with multipartite systems.
Suppose for simplicity to have a tripartite spin system $S_{123}=S_1 \times S_2 \times S_3$, 
in this case we must specify what kind of entanglement we mean.

Indeed, one can consider bipartite entanglement for each one of the three possible bipartitions of the system, namely
$S_1 \times S_{23}$, $S_2 \times S_{31}$ and $S_3 \times S_{12}$, otherwise one could be interested in tripartite 
entanglement: this yields to the classification of states discussed in \cite{cirac}.

We say that a state is tri-separable if density matrix admits a decomposition as a convex sum of local states \cite{Werner}:
\begin{equation}
\rho = \sum_jp_j \rho_1^{(j)} \otimes \rho_2^{(j)} \otimes \rho_3^{(j)},\qquad \sum_jp_j=1,\qquad p_j \geq 0
\end{equation}
otherwise the state is entangled.

The following relation holds between bi-separability and tri-separability:
if a state is tri-separable it is also bi-separable for each choice of the bipartition, 
but a state can be bi-separable for certain bipartion but not tri-separable.
It was shown in \cite{Bennet} that there are also states that are bi-separable for all possible bipartition of the 
system but are tri-entangled.

It is clear that a map like (\ref{wit}) cannot be used in order to obtain tri-entanglement witness because it is 
sensitive only to bipartite entanglement.
In order to investigate tripartite separability one has to consider maps like the following:
\begin{equation}
\mathcal{P}=I \otimes P \otimes P'
\end{equation}
where $P$ and $P'$ are positive but not completely positive maps acting on the second and the third subsystem respectively.

It is clear that it cannot be sufficient to consider only transposition in order to obtain tripartite entanglement witness, 
this is simply because $I \otimes T_2 \otimes T_3 = I \otimes T_{23}$ and one can obtain only a bipartite entanglement 
witness.
Thus one can consider at least two different positive but not completely positive maps.

Let us suppose that we have three families of positive but not completely positive maps acting on the subsystems
$P_i(\lambda_i)$ for $i=1,2,3$ depending on an array of parameters that we indicate with the short hand notation 
$\lambda_i$, and consider the map acting on the whole system
\begin{equation}
\mathcal{P}_{\lambda_1 \lambda_2 \lambda_3}=P_1(\lambda_1) \otimes P_2(\lambda_2) \otimes P_3(\lambda_3)
\end{equation}

Let us also suppose that for a certain value of the parameters, say $\lambda_i=1$, $P_i(0)=I_i$ thus we can study 
tripartite entanglement by varing all the parameters or bipartite entanglement by fixing two of them to zero.

\section{\label{Scaling transform}Scaling transform}

In a previous paper \cite{Anil} a family of maps was presented that can be useful in order to detect entanglement in
multimode pure Gaussian states \cite{Simon,Anil} of electromagnetic field.
It is defined in a continous variables system as a rescaling of momentum:
\begin{eqnarray} \label{scale}
\left\{ \begin{array}{ccc}
x & \longrightarrow & x         \\
p & \longrightarrow & \lambda p  
\end{array}\right.
\end{eqnarray}

For $\lambda \in [-1,1]$ it defines a semigroup of maps that are non canonical almost everywhere.
The map (\ref{scale}) is physically equivalent to a rescaling of time $t \longrightarrow \lambda t$ or Planck's constant
$\hbar \longrightarrow \lambda \hbar$.
In the case of $\lambda=-1$ it reduces to time reversal while for $\lambda=1$ is the identical map.
The fact that (\ref{scale}) is in general non canonical play a foundamental role in order to detect whether a given 
state is entangled \cite{Anil}.
It the present letter we propose a realization of this map in the case of spin systems, that is, those
quantum systems with discrete variables.
In the case of a qubit a generic (mixed) state is written as
\begin{eqnarray}
\rho=\frac{1}{2}\left[\begin{array}{cc}
1+z & x-iy \\
x+iy & 1-z
\end{array}\right]
\end{eqnarray}
One can define:
\begin{eqnarray}
T_\lambda \ \ | \ \ \rho \rightarrow T_\lambda \rho =
\frac{1}{2}\left[\begin{array}{cc}
1+z          & x-i\lambda y \\
x+i\lambda y & 1-z
\end{array}\right]
\end{eqnarray}

$T_\lambda$ defines a scaling transformation that for $\lambda=1$ is the identical map and for $\lambda=-1$ reduces
to the transposition or time reversal.

The matrix representation of the map is
\begin{eqnarray} T_\lambda \equiv
\left[\begin{array}{cccc}
1 & 0                      & 0                      & 0 \\
0 & \frac{1}{2}(1+\lambda) & \frac{1}{2}(1-\lambda) & 0 \\
0 & \frac{1}{2}(1-\lambda) & \frac{1}{2}(1+\lambda) & 0 \\
0 & 0                      & 0                      & 1
\end{array}\right].
\end{eqnarray}
that is, the matrix $T_\lambda$ transforms the vector
\begin{eqnarray} \vec\rho=
\left[\begin{array}{c}
\rho_{11}\\\rho_{12}\\\rho_{21}\\\rho_{22}
\end{array}\right]
\end{eqnarray}
which corresponds to qubit density matrix
\begin{eqnarray} \rho=
\left[\begin{array}{cc}
\rho_{11}&\rho_{12}\\
\rho_{21}&\rho_{22}
\end{array}\right]
\end{eqnarray}
into a new vector
\begin{eqnarray} \vec\rho_\lambda=
\left[\begin{array}{c}
\rho_{11}\\\frac{1+\lambda}{2}\rho_{12}+\frac{1-\lambda}{2}\rho_{21}\\
\frac{1-\lambda}{2}\rho_{12}+\frac{1+\lambda}{2}\rho_{21}\\
\rho_{22}
\end{array}\right].
\end{eqnarray}
This new vector corresponds to new density matrix
\begin{eqnarray} \rho_\lambda=
\left[\begin{array}{cc}
\rho_{11}&\frac{1+\lambda}{2}\rho_{12}+\frac{1-\lambda}{2}\rho_{21}\\
\frac{1-\lambda}{2}\rho_{12}+\frac{1+\lambda}{2}\rho_{21}&\rho_{22}
\end{array}\right].
\end{eqnarray}

In order to determine if this map is completely positive one has to consider the auxiliary matrix:
\begin{eqnarray} B_\lambda =
\left[\begin{array}{cccc}
1                      & 0                      & 0                      & \frac{1}{2}(1+\lambda)  \\
0                      & 0                      & \frac{1}{2}(1-\lambda) & 0                       \\
0                      & \frac{1}{2}(1-\lambda) & 0                      & 0                       \\
\frac{1}{2}(1+\lambda) & 0                      & 0                      & 1
\end{array}\right].
\end{eqnarray}
The map is completely positive if and only if the matrix $B_\lambda$ is positive\cite{Sudar61,Zyc}, and it is so 
only for $\lambda=1$.
In analogy to the fact that (\ref{scale}) is non canonical, $T_\lambda$ is neither unitary nor completely positive
for almost all values of $\lambda$.

Notice that $T_\lambda$ can be written as a convex sum of the identity map and transposition:
\begin{equation} \label{deff}
T_\lambda = \frac{1+\lambda}{2}I + \frac{1-\lambda}{2}T
\end{equation}
Being a sum of a completely positive map and transposition, it is clear that $T_\lambda$ cannot be useful in order 
to detect bound entanglement, i.e. $T_\lambda$ can detect entanglement only if $T$ can.
As a consequence bound entangled states cannot be detect by the map $T_\lambda$.
Equation (\ref{deff}) can be taken as a definition of the scaling transformation for a qu$d$it system.

\section{\label{examples}Some examples}

In this section we are going to test the power of the scaling tranform to detect entanglement in bipartite and tripartite 
systems.

\subsection{Werner states for two qubits}

In this section we are going to test the capability of the maps $T_\lambda$ as sources of entanglement witnesses 
in the case of a two qubit system.
In order to do that we apply the one parameter family of maps:
\begin{equation} \label{FIG1lambda}
\mathcal{T}_\lambda=I \otimes T_\lambda
\end{equation}
to the well known Werner states \cite{Werner}.
Werner states for two qubits are defined as follows:
\begin{equation} \label{introw}
w_p = p|\Psi\rangle\langle\Psi| + (1-p)\frac{I_{4}}{4}
\end{equation}
where $|\Psi\rangle$ is a maximally entangled (Bell) state:
\begin{equation}
|\Psi\rangle = \frac{1}{\sqrt{2}} \left( |00\rangle + |11\rangle \right)
\end{equation}
and $I_4/4$ is the maximally mixed state for a two qubits system.
The density matrix expression for a Werner state is:
\begin{eqnarray} \label{Wmat}
w_p = \left[\begin{array}{cccc}
\frac{1+p}{4} & 0             & 0             & \frac{p}{2} \\
0             & \frac{1-p}{4} & 0             & 0           \\
0             & 0             & \frac{1-p}{4} & 0           \\
\frac{p}{2}   & 0             & 0             & \frac{1+p}{4}
\end{array}\right]
\end{eqnarray}

The matrix (\ref{Wmat}) is known to be positive for all $p \in [-\frac{1}{3} , 1]$, separable for 
$p \in [-\frac{1}{3} , \frac{1}{3}]$ and entangled for $p \in (\frac{1}{3} , 1]$.

Applying the $1 \otimes T_\lambda$ map one gets:
\begin{eqnarray} \label{FG1}
w_{p;\lambda} = \mathcal{T}_\lambda w_p = \left[ \begin{array}{cccc}
\frac{1+p}{4}                               & 0                                           & 0                                           & \frac{p}{2}\left(\frac{1+\lambda}{2}\right) \\
0                                           & \frac{1-p}{4}                               & \frac{p}{2}\left(\frac{1-\lambda}{2}\right) & 0                                           \\
0                                           & \frac{p}{2}\left(\frac{1-\lambda}{2}\right) & \frac{1-p}{4}                               & 0                                           \\
\frac{p}{2}\left(\frac{1+\lambda}{2}\right) & 0                                           & 0                                           & \frac{1+p}{4}
\end{array}\right]
\end{eqnarray}
with $\lambda \in [-1 , 1]$.

The four eigenvalues of the matrices $w_{p;\lambda}$ are:
\begin{eqnarray}
\epsilon_1(p;\lambda)=\epsilon_2(p;\lambda) &=& \frac{1-p\lambda}{4}    \\
\epsilon_3(p;\lambda)                       &=& \frac{1-2p+p\lambda}{4} \\
\epsilon_4(p;\lambda)                       &=& \frac{1+2p+p\lambda}{4}
\end{eqnarray}
The map $\mathcal{T}_\lambda$ detects entanglement in Werner states if some of this eigenvalues is strictly negative.

A simple calculation shows that $T_\lambda$ detects negativity of matrix $w_{p;\lambda}$ only if
$p \geq (2-\lambda)^{-1}$ hence $T_\lambda$ gives a necessary and sufficient condition for separability since for 
$\lambda = -1$ all the Werner states with $p>1/3$ give at least one negative eigenvalue.

\begin{figure}\centering
\includegraphics[width=0.5\textwidth, height=0.25\textwidth]{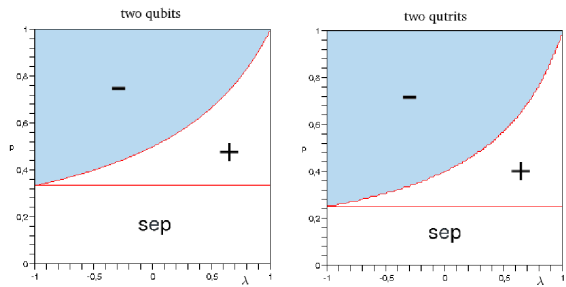}
\caption{on the left: phase diagram for matrices (\ref{FG1}): on the vertical axis the parameter $p$ 
definig the state (\ref{introw}) is plotted, on the horizontal axis the parameter $\lambda$ 
definig the map (\ref{FIG1lambda}) is plotted. The "sep" indicates the domain of separability of
(\ref{Wmat}), the coloured regions corresponds to negative eigenvalues in (\ref{FG1}) that witness entanglement.
On the right: the anologous phase diagram for two qutrits state (\ref{q3}).}
\label{2qubits}
\end{figure}

\subsection{Measure of entanglement}

In \cite{SudZac} the unitary spin tomogram of qubit state was introduced.
It is defined as joint probability distribution $w$ of spin projections $m_1,m_2$ depending on unitary matrix elements 
$u\in U(4)$ group.
The tomogram of a state of two qubits with density matrix $\rho$ reads
\begin{equation}
w_\rho(m_1,m_2,u)=\langle m_1,m_2\mid u^\dagger\rho u\mid m_1,m_2\rangle.
\end{equation}
The tomogram completely determines the density matrix $\rho$~\cite{Olga,SudZac}.
The measure of entanglement introduced in \cite{SudZac} reads
\begin{equation}
{\cal M}(\rho)=\max\sum_{m_1m_2}\Big(|w_{\rho_M}(m_1,m_2,u)|-w_{\rho_M}(m_1,m_2,u)\Big).
\end{equation}
The function ${\cal M}(\rho)=0$ for separable states and it is positive for entangled states.
The maximum is taken with respect to all unitary group elements and with respect to all positive but not completely 
positive maps $M$ of subsystem density matrix which indices the map of composed system density matrix 
$\rho\rightarrow\rho_M$.

If one applies all the maps one will have for some of them the matrix $\rho_M$ to be nonpositive for entangled state $\rho$.

Below we applied the maps which depend on parameters $\lambda$.
In view of this, the maximum with respect to unitary matrix elements can be achieved if one takes the eigenvalues 
of the matrix $\rho_M$.
The maximum with respect to all positive but not completely positive maps $M$ of the subsystem density matrix becomes 
the maximum with respect to $\lambda$.

Below we consider the eigenvalues of the matrix $\rho_M$ and apply the constructed measure in the reduced
form.

For the trace preservation property the sum of the eigenvalues equals one for all values of $\lambda$:
\begin{equation}
\sum_{j=1}^4 \epsilon_j(p;\lambda) = 1
\end{equation}

We can introduce the quantity:
\begin{equation}
m(p,\lambda) = \sum_{j=1}^4 |\epsilon_j(p,\lambda)| - 1
\end{equation}
that is zero when all the eigenvalues are positive and is strictly positive when at least one eigenvalue becomes negative.
If we take the maximum over all value of $\lambda \in [-1,1]$
\begin{equation}
M(p)=max_{\lambda \in [-1,1]} \{ m(p,\lambda)\}
\end{equation}
this defines a measure of entanglement.

For the previous example this maximum is reached at $\lambda=-1$ for all values of $p$ and yields:
\begin{eqnarray}
M(p)= \left\{ \begin{array}{ccc}
0              & \mbox{for} & p \leq \frac{1}{3} \\
\frac{3p-1}{2} & \mbox{for} & p > \frac{1}{3}
\end{array}\right.
\end{eqnarray}

Here it coincides with negativity measure \cite{Horod}.

\subsection{States of three qubits}

In this section we investigate the more interesting case of tripartite system, that is, a system of three qubits.

For a composite system of $n$ qu-$d$its one can consider the states:
\begin{equation} \label{wer}
w'_p = p | w' \rangle\langle w' | + (1-p)\frac{I_{d^n}}{d^n}
\end{equation}
where $|w'\rangle$ is a maximally entangled Greenberger-Horne-Zeilinger like \cite{GHZ} state:
\begin{equation} \label{GG}
|w'\rangle = \frac{1}{\sqrt{d}} \sum_{i=0}^{d-1} |i\rangle_1 \otimes |i\rangle_2 \otimes ... |i\rangle_n
\end{equation}
The states (\ref{wer}) are known \cite{werner} to be entangled for $p \geq p_{ent}$ and separable for $p \leq p_{ent}$,
with
\begin{equation} \label{pent}
p_{ent} = \frac{1}{d^{n-1}+1}
\end{equation}

For a system of three qubits (\ref{GG}) is the $GHZ$ state:
\begin{equation}
|GHZ\rangle = \frac{1}{\sqrt{3}}\left( |000\rangle + |111\rangle \right)
\end{equation}
this yields the following density matrix:
\begin{eqnarray}
w'_p = \left[\begin{array}{cccccccc}
\frac{p}{2}+\frac{1-p}{8} & 0             & 0             & 0             & 0             & 0             & 0             & \frac{p}{2}              \\
0                         & \frac{1-p}{8} & 0             & 0             & 0             & 0             & 0             & 0                        \\
0                         & 0             & \frac{1-p}{8} & 0             & 0             & 0             & 0             & 0                        \\
0                         & 0             & 0             & \frac{1-p}{8} & 0             & 0             & 0             & 0                        \\
0                         & 0             & 0             & 0             & \frac{1-p}{8} & 0             & 0             & 0                        \\
0                         & 0             & 0             & 0             & 0             & \frac{1-p}{8} & 0             & 0                        \\
0                         & 0             & 0             & 0             & 0             & 0             & \frac{1-p}{8} & 0                        \\
\frac{p}{2}               & 0             & 0             & 0             & 0             & 0             & 0             & \frac{p}{2}+\frac{1-p}{8}
\end{array}\right]
\end{eqnarray}

Because of the particular symmetric form of the states (\ref{wer}) they are bi-entangled if and only if tri-entangled.
In order to appreciate separately bipartite entanglement and tripartite entanglement we modify states (\ref{wer}) 
introducing an addictional parameter:
\begin{equation} \label{ptheta}
w'_{p,\theta} = p |\Psi_\theta\rangle \langle\Psi_\theta| + (1-p)\frac{I_8}{8}
\end{equation}
for $\theta \in [0,\frac{\pi}{2}]$, where
\begin{equation} \label{theta3}
|\Psi_\theta\rangle = \frac{1}{\sqrt{2}}\left( |000\rangle + \cos{\theta}|111\rangle + \sin{\theta}|110\rangle \right)
\end{equation}

These states have no more the symmetric form of (\ref{wer}) and we expect a different behaviour when describing
tripartite and bipartite entanglement with respect to different bipartition of the system.

The pure state $|\Psi_\theta\rangle\langle\Psi_\theta|$ contributes to the entanglement of (\ref{ptheta}) and
the identity tends to remove purity as well entanglement while $p$ is decreasing from $1$ to $0$.

The state vector in (\ref{theta3}) is maximally entangled for $\theta=0$ for all possible choices of partitions of the 
system, while for $\theta=\pi/2$ it is separable for the decompositions $S_{12} \times S_3$ while it is still 
maximally entangled for the decomposition $S_1 \times S_{23}$ and $S_2 \times S_{31}$.

Notice that the states $w'_{p,\theta}$ in (\ref{ptheta}) are symmetric for the interchange of the first and the second 
system: this implies that we can consider maps with two parameters only:
\begin{equation} \label{3qubitsmap}
I \otimes T_\lambda \otimes T_\mu
\end{equation}
for $\lambda=1$ we can study separability with respect to the bipartition $S_{12} \times S_3$, for $\mu=1$ the 
bipartition $S_{31} \times S_2$ (that is the same as $S_1 \times S_{23}$) and by varing both $\lambda$ and $\mu$ we can 
study tripartite entanglement.

We consider the matrices:
\begin{equation} \label{pthetalambdamu}
w'_{p,\theta;\lambda,\mu} = (I \otimes T_\lambda \otimes T_\mu) w'_{p,\theta}
\end{equation}
$w'_{p,\theta}$ is surely entangled if there is at least one value of $\lambda$ and $\mu$ for which 
$w'_{p,\theta;\lambda,\mu}$ is strictly non-positive.

More precisely if there is some negative eigenvalue on the line $\lambda=1$ this witness a bipartite entanglement with 
respect to the decomposition $S_{12} \times S_3$, if a negative eigenvalue is found on the line $\mu=1$ this witness a 
bipartite entanglement with respect to the decomposition $S_{31} \times S_2$, finally a negative eigenvalue in generic 
point witnesses tripartite entanglement, while the point $\lambda=\mu=-1$ gives informations on the bipartition 
$S_1 \times S_{23}$.

In the symmetric case $\theta=0$ the eigenvalues, as functions of $p$, $\lambda$ and $\mu$, are:
\begin{eqnarray}
\epsilon_1 &=& \frac{p}{2} + \frac{1-p}{8} + \frac{p}{8}\left(1+\lambda\right)\left(1+\mu\right) \\
\epsilon_2 &=& \frac{p}{2} + \frac{1-p}{8} - \frac{p}{8}\left(1+\lambda\right)\left(1+\mu\right) \\
\epsilon_3 &=& \frac{1-p}{8} + \frac{p}{8}\left(1+\lambda\right)\left(1-\mu\right)               \\
\epsilon_4 &=& \frac{1-p}{8} - \frac{p}{8}\left(1+\lambda\right)\left(1-\mu\right)               \\
\epsilon_5 &=& \frac{1-p}{8} + \frac{p}{8}\left(1-\lambda\right)\left(1+\mu\right)               \\
\epsilon_6 &=& \frac{1-p}{8} - \frac{p}{8}\left(1-\lambda\right)\left(1+\mu\right)               \\
\epsilon_7 &=& \frac{1-p}{8} + \frac{p}{8}\left(1-\lambda\right)\left(1-\mu\right)               \\
\epsilon_8 &=& \frac{1-p}{8} - \frac{p}{8}\left(1-\lambda\right)\left(1-\mu\right)
\end{eqnarray}

In order to measure entanglement we can consider the function
\begin{equation}
m(p;\lambda,\mu) = \sum_{i=1}^8 |\epsilon_i(p,0;\lambda,\mu)| -1
\end{equation}
this function is zero when the matrix $w'_{p;\lambda,\mu}$ is positive and is strictly greater than zero when at least 
one eigenvalue $\epsilon_i$ is negative.

A measure of entanglement is obtained by taking the maximum over $\lambda$ and $\mu$:
\begin{equation}
M(p) = max_{\lambda,\mu}\{ m(p;\lambda,\mu) \}
\end{equation}
it is easy to show that this maximum is reached at point $\lambda=-1$ and $\mu=-1$ and that 
$m(p;-1,-1)=m(p;-1,1)=m(p;1,-1)$, this yields:
\begin{eqnarray}
M(p) = \left\{
\begin{array}{ccc}
0              & \mbox{for} & -\frac{1}{7} \leq p \leq \frac{1}{5} \\
\frac{5p-1}{4} & \mbox{for} &  \frac{1}{5} < p \leq 1
\end{array} \right.
\end{eqnarray}
That is, the amount of bipartite and tripartite entanglement is the same.

\begin{figure}\centering
\includegraphics[width=0.5\textwidth, height=0.75\textwidth]{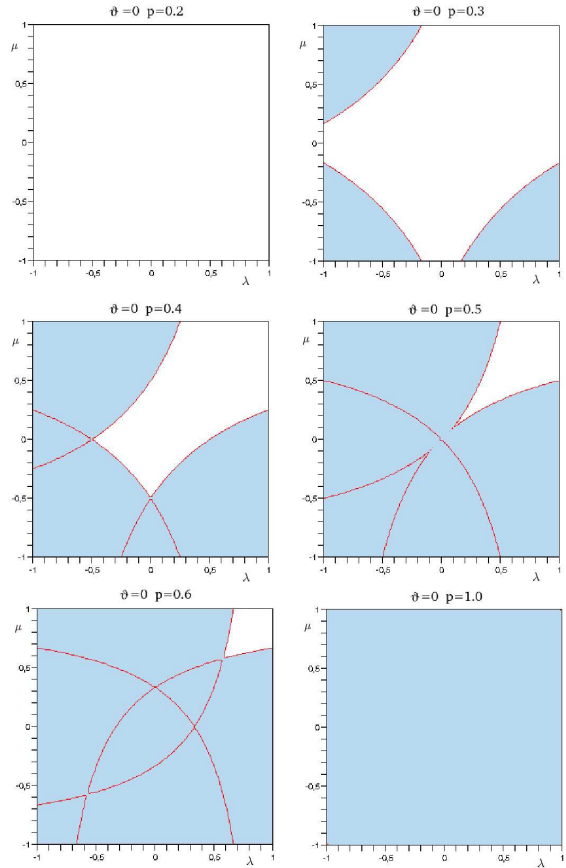}
\caption{three qubit state (\ref{ptheta}): some phase diagrams for $\theta=0$ for several values of $p$.
On the horizontal and vertical axis the two parameters $\lambda$ and $\mu$ defining the map (\ref{3qubitsmap}) are plotted.
Coloured regions indicate negative eingenvalues of (\ref{pthetalambdamu}) that witness entanglement of state (\ref{ptheta}).
For $p \leq 1/5$ the state is fully separable.}
\label{GHZ0}
\end{figure}

\begin{figure}\centering
\includegraphics[width=0.5\textwidth, height=0.75\textwidth]{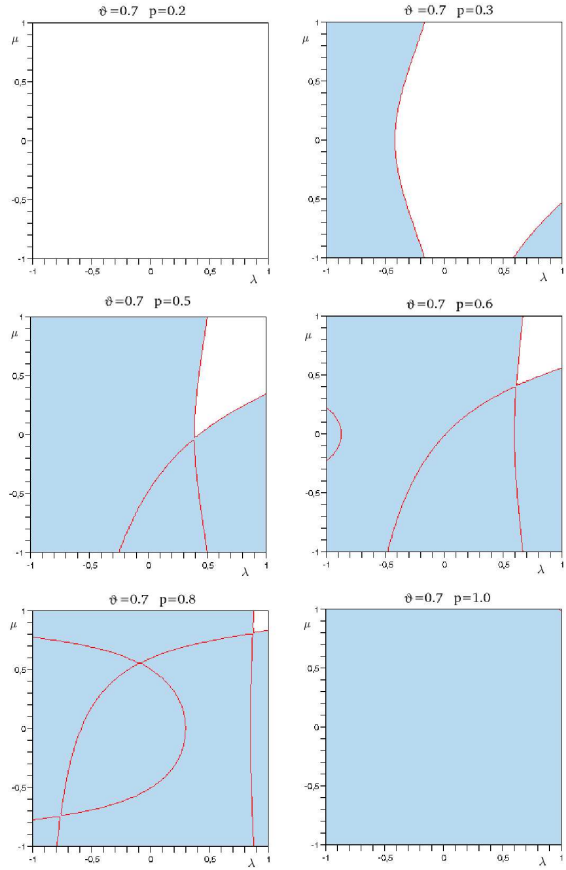}
\caption{three qubit state (\ref{ptheta}): some phase diagrams for $\theta=0.7$.
On the axis the map parameters (\ref{3qubitsmap}) $\lambda$ and $\mu$ are plotted. 
The states are entangled in any sense for $p>1/5$.}
\label{GHZ1}
\end{figure}

\begin{figure}\centering
\includegraphics[width=0.5\textwidth, height=0.5\textwidth]{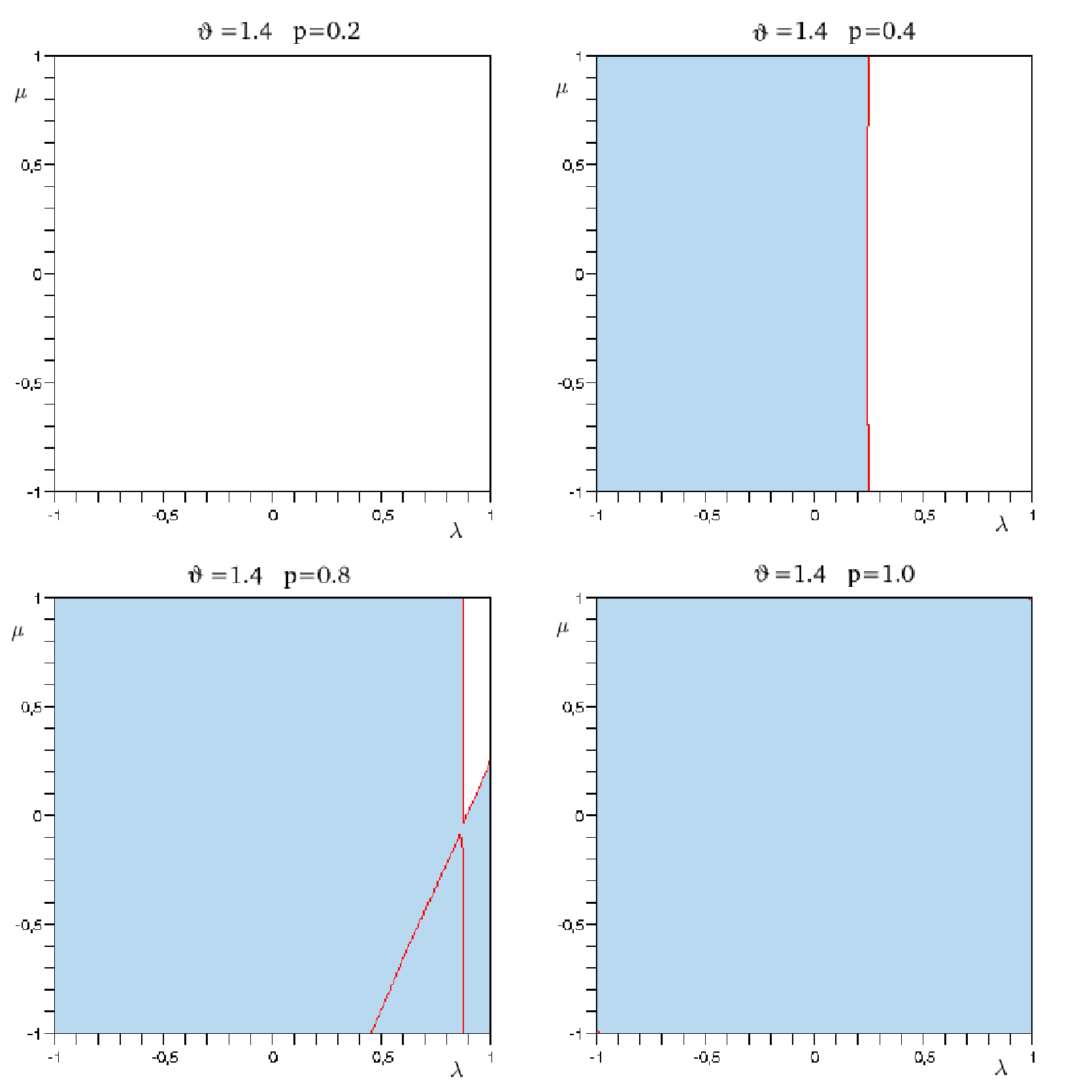}
\caption{three qubit state (\ref{ptheta}): some phase diagrams for $\theta=1.4$.
On the axis the map parameters (\ref{3qubitsmap}) $\lambda$ and $\mu$ are plotted.
The state could be separable for $p \leq 1/5$.
For $p=0.4$ it is shown that the state is tri-entangled and bi-entangled for 
$S_1 \times S_{23}$ and $S_2 \times S_{31}$ but it could be bi-separable for $S_3 \times S_{12}$.}
\label{GHZ2}
\end{figure}

The general case $\theta \neq 0$ can be worked by numerical analysis and some results are shown in figures 
(\ref{GHZ1},\ref{GHZ2}).

For example for $\theta=1.4$ and $p=0.4$ the state is surely tri-entangled and bi-entangled for the bipartitions
$S_2 \times S_{13}$ and $S_1 \times S_{23}$, but it could be separable for the bipartition $S_3 \times S_{12}$.

\subsection{W states and qutrits}

In this section we study the states:
\begin{equation} \label{pWstate}
p|W \rangle\langle W| + \frac{1-p}{8}I_8
\end{equation}
where $|W\rangle$ is the $W$ state \cite{Wstates}:
\begin{equation}
|W\rangle = \frac{1}{\sqrt{3}}\left( |001\rangle + |010\rangle + |100\rangle \right)
\end{equation}

\begin{figure}\centering
\includegraphics[width=0.5\textwidth, height=0.5\textwidth]{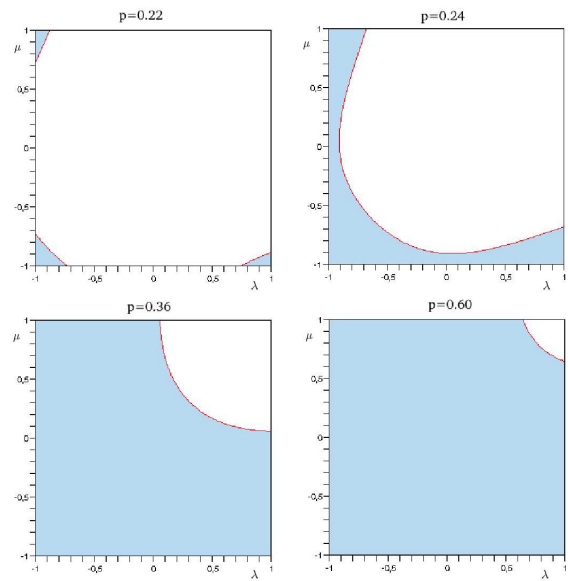}
\caption{three qubits state (\ref{pWstate}): some phase diagrams for several values of the state parameters 
$p$ (\ref{pWstate}). On the axis the map parameters (\ref{3qubitsmap}) $\lambda$ and $\mu$ are plotted.}
\label{W1}
\end{figure}

The amount of tripartite entanglement is equal to that of bipartite entanglement for all the possible bipartitions:
as an example the tripartite entanglement measure for $p=0.6$ is
\begin{equation}
M(p=0.6) = m(0.6,-1,-1)=m(0.6,-1,1)=m(0.6,-1,1)\cong 1.47
\end{equation}

As a simple example for a system of two qutrits we consider the $GHZ$-like state
\begin{equation} \label{q3}
p|\psi \rangle\langle \psi| + \frac{1-p}{p}I_9
\end{equation}
where now
\begin{equation}
|\psi\rangle = \frac{1}{\sqrt{3}}\left( |00\rangle + |11\rangle + |22\rangle  \right)
\end{equation}

As in (\ref{introw}) the states in (\ref{q3}) refer to a bipartite system, hence we need only one parameter: 
see the right side of figure (\ref{2qubits}).
For (\ref{pent}) the state is fully separable for $p \leq 1/4$.

\section{\label{outlook}Outlook}

We have seen that the proposed scaling transform cannot be useful to detect bound entanglement, nevertheless it is 
useful to study multipartite entanglement.
In future publication we are going to consider a larger family of maps increasing the number of parameters.
It is known that for qubits the set of positive maps is described by 12 paramenters \cite{Zyc}.
One could ask if it is possible to choice $n < 12$ paramenters that are sufficient in order to witness the entanglement 
of any non-separable states.

Intuitively one can consider partial scaling of time transform with changing of "rotation angular velocity" which corresponds to
changes
$$\sigma_x\rightarrow\mu_1\sigma_x,\quad\sigma_y\rightarrow\mu_2\sigma_y,\quad
\sigma_z\rightarrow\mu_3\sigma_z$$ for qubits. Analogous transform can be obtained for orbital momentum.

For $\mu_1,\mu_2,\mu_3 \in [-1,1]$ this is a three parameter semigroup of transformations that corresponds to non
isotropic contraction of the Bloch ball.
These maps are positive for all the values of the parameters but are completely positive only for points inside
a tetrahedron in the space of parameters with vertices $(1,1,1);(1,-1,-1);(-1,1,-1);(-1,-1,1)$ \cite{Zyc}.

\section{\label{conclusions}Conclusions}

To summarize the results obtained, we can point out that we have suggested a new criterion of separability for
multiqubit state. 
The criterion is based on partial scaling transform of time (or Planck's constant) which provides positive but not 
completely positive map of qubit density matrix.
Using the partial scaling map of density matrix of composite multiqubit system, one can detect the entanglement.
The partial scaling criterion is reduced to partial transpose criterion for particular scaling parameter equal $-1$.

We used the tomographic measure of entanglement which for the value of scaling parameter equal $-1$ coincides
with negativity measure.

The Werner states for two qubit system and some families of states for tripartite system and bipartite qutrits system 
were investigated and the values of the state parameters for which the system state is entangled were determined using 
the suggested criterion.

In future publication, we study other examples of multiqubit states using the positive partial scaling transform
criterion with increased number of parameters towards a complete family of separability criteria.

\ack{V.I.M and E.C.G.S. thank Dipartimento di Scienze Fisiche, 
Universit\`a di Napoli Federico II and Istituto Nazionale di Fisica Nucleare 
Sezione di Napoli for kind hospitality.}

\end{document}